\documentclass[preprint2]{aastex}
\usepackage{graphicx}
\usepackage{natbib}

\shorttitle{Optical Variability in the GOODS-South Field}
\shortauthors{Klesman $\&$ Sarajedini}

\begin{document}

\title{Optical Variability of Infrared Power Law-Selected Galaxies \\ $\&$ X-ray Sources in the GOODS-South Field}

\author{Alison Klesman $\&$ Vicki Sarajedini}
\affil{Department of Astronomy, University of Florida, Gainesville, FL 32611}

\email{alichan@astro.ufl.edu}
\email{vicki@astro.ufl.edu}

\begin{abstract}
We investigate the use of optical variability to identify and study Active Galactic Nuclei (AGN) in the GOODS-South field. A sample of 22 mid-infrared power law sources and 102 X-ray sources with optical counterparts in the HST ACS images were selected. Each object is classified with a variability significance value related to the standard deviation of its magnitude in five epochs separated by 45-day intervals. The variability significance is compared to the optical, mid-IR, and X-ray properties of the sources. We find that 26$\%$ of all AGN candidates (either X-ray- or mid-IR-selected) are optical variables. The fraction of optical variables increases to 51$\%$ when considering sources with soft X-ray band ratios. For the mid-IR AGN candidates which have multiwavelength SEDs, we find optical variability for 64$\%$ of those classified with SEDs like Broad Line AGNs. While mostly unobscured AGN appear to have the most significant optical variability, some of the more obscured AGNs are also observed as variables. In particular, we find two mid-IR power law-selected AGN candidates without X-ray emission that display optical variability, confirming their AGN nature.  
\end{abstract}

\keywords{galaxies: active}

\section{INTRODUCTION}

Active Galactic Nuclei (AGN) are envisioned to be accreting supermassive black holes at the centers of galaxies. Monitoring variability in these sources is an important way to extract information about the compact object within, as the timescales over which this variability is seen can provide information about the size and structure of the region from which the emission originates in order to place physical constraints on the central engine within. AGN are known to show variability on a variety of timescales: the optical line emission and continuum flux from quasars (QSOs) has been observed to vary on timescales of months to years, while X-ray flux from these sources varies on shorter timescales of hours to days (Peterson 2001 $\&$ references therein). The mechanism behind this variability is still uncertain, though many theories have been formulated to explain this phenomenon. The leading explanation for the observed variability is that changes in magnitude are related to accretion events or disk instabilities \citep{per06}. Other explanations for variability include changes in magnitude due to supernovae, starbursts, or microlensing events (e.g., Aretxaga $\&$ Terlevich 1994).

It has been shown that 80-100\% of AGN candidates vary in the optical over the course of several years (e.g., Koo, Kron, $\&$ Cudworth 1986). Webb \& Malkan (2000) found that 60\% of Low-Luminosity AGNs (LLAGNs, those with luminosities M$_{B}$$\ge$-23) varied when studied over a period of months. Thus, optical variability is an important and successful method for identifying AGN in imaging surveys. Indeed, there is some evidence that LLAGN exhibit greater amplitude flux changes than do their brighter counterparts \citep{ber98}, making variability a particularly effective means for their identification. LLAGN represent a significant contribution to the X-ray, IR, and UV backgrounds, so knowledge of their space and luminosity distribution at a range of redshifts is important. They may also represent an important phase of galaxy evolution, which connects normal and active galaxies \citep{cro05}.

This paper presents the initial results from a variability survey of the Great Observatories Origins Deep Survey (GOODS) South and North fields to identify optically varying AGN, particularly LLAGN, at z$\sim$1 using small aperture photometry to isolate galactic nuclei. This technique has been shown to successfully identify faint AGN in high-resolution HST multi-epoch imaging surveys \citep{sgk03, sar06}.

The GOODS images were obtained with the HST Advanced Camera for Surveys (ACS) over five epochs separated by 45-day intervals, allowing for optical variability to be monitored over a six-month period. Our initial study focuses on a pre-selected sample of galaxies in the GOODS-South field based on X-ray and/or IR properties. In this paper, we investigate the optical variability of AGN candidates having a range of obscuring properties, from soft X-ray sources to highly obscured mid-IR sources lacking X-ray emission. The aim of our study is twofold: 1) to confirm via optical variability the AGN nature of candidates identified in multiwavelength surveys, and 2) to quantify the use of optical variability in identifying various types of AGN for use in interpreting the results of our larger forthcoming survey of all optical sources in the GOODS South and North fields.

\section{AGN CANDIDATES}

The galaxies in this study were chosen from two catalogs. The first is a sample of mid-infrared-selected AGN candidates from Alonso-Herrero et al. (2006) with optical counterparts in the GOODS-South field. These galaxies were selected using Spitzer/MIPS 24 $\mu$m observations and are well-fit with a power law spectral energy distribution (SED) through the Spitzer IRAC bands at 3.6, 4.5, 5.8, and 8 $\mu$m. The second sample consists of X-ray sources compiled by Giacconi et al. (2002) and Alexander et al. (2003) using the Chandra X-ray observatory. The Chandra Deep Field South (CDFS), which encompasses the GOODS-South ACS field, is a combination of 11 individual pointings taken with the Advanced Camera Imaging Spectrometer (ACIS-I) covering a range of 0.2-10 keV with a total exposure time of 942 ks. 

In the next sections we discuss how the sources in these catalogs were cross-matched with optical sources in the ACS fields for inclusion in our variability survey. It was first necessary to estimate the optical magnitude to which photometry could be accurately and robustly obtained. To do this, we obtained photometry of optical sources in four epoch 1 frames (sections 4, 8, 10, and 12) to cover a large spatial range in the GOODS field. We used the IRAF tool DAOfind to locate all sources in order to construct a histogram of number vs. nuclear (r = 0.075" or 2.5 pixels) V magnitude. We measured the magnitude of the galaxy nuclei as we would be doing to identify nuclear variability in the selected samples. This histogram revealed a sudden drop in number counts beginning at a nuclear V magnitude of 27. We therefore chose a magnitude cutoff of 27 and any objects with average magnitudes fainter than V$_{nuc}$ = 27 were not considered in our analysis. This limit is also consistent with the limiting magnitude to which nuclear variability may expect to be detected using fixed aperture photometry for galactic nuclei extending to z$\sim$1. In the current dataset, we expect to become less sensitive to a varying nucleus within a host galaxy at nuclear V magnitudes of $\sim$27.5. \footnote{See discussion in Sarajedini, Gilliland, $\&$ Kasm (2003), Section 3, for further discussion of this effect.} 

\subsection{IR-Selected Galaxies}

Alonso-Herrero et al. (2006) identified 92 galaxies within the CDFS which were detected at 24 $\mu$m and display power law-like emission from 3.6 to 8 $\mu$m. These galaxies were selected at 24 $\mu$m because at z$>$1 a large fraction of these objects should be Ultra-Luminous InfraRed Galaxies (ULIRGs, with L$_{IR}$$>$10$^{12}$ L$_{sun}$). ULIRGs are known to have a steep power law-like SED in the infrared \citep{san88,kla01} and many are classified as QSOs. Alexander et al. (2005) showed that at least 40\% of IR-luminous high-redshift galaxies contain AGN, and similar results at z=1-2 have been found with recent Spitzer observations \citep{yan05}. All of the sample galaxies are luminous in the IR: 30\% are hyper luminous (L$_{IR}$$>$10$^{13}$  L$_{sun}$), 41\% are ULIRGs (L$_{IR}$$\sim$10$^{12}$ - 10$^{13}$ L$_{sun}$), and all but one are Luminous IR Galaxies (LIRGs, L$_{IR}$$\sim$10$^{11}$ - 10$^{12}$ L$_{sun}$). Alonso-Herrero et al. (2006) have further quantified the multiwavelength SEDs to classify their sources as either Broad Line AGNs (BLAGNs) or Narrow Line AGNs (NLAGNs). 

We matched sources identified in the ACS GOODS field from the version 1.1 source catalog derived from the stacked V-band image with the IR catalog and found 37 matches within a 3.5" radius, 30 of which fell within 1". Of these matches, 11 were too faint for optical photometry (fainter than 27th magnitude in a single V-band epoch) and 5 were not visible in all five GOODS epochs. This left a total of 22 targets. All final optical sources matched the IR coordinates within 1.1". 

\subsection{X-ray-Selected Galaxies}

AGN as a class are known to be X-ray sources and are commonly selected via X-ray surveys \citep{bh05}. We matched the coordinates of objects in the Alexander et al. (2003) Chandra X-ray catalog with the GOODS version 1.1 source catalog and found 200 matches within a 1.9" radius. Of these optical counterparts, 80 were too faint for optical photometry, 16 were not visible in all five epochs, one was contaminated by a cosmic ray, and one optical source was added due to a double nucleus. Finally, one additional source was found by matching the GOODS catalog coordinates with the Giacconi et al. (2002) X-ray catalog, which also contains the 102 X-ray sources from Alexander et al. Thus, a total of 104 X-ray-selected objects were photometered to look for optical variability. Fourteen of these objects overlap the IR-selected sources discussed in Section 2.1.

\section{ACS GOODS DATA $\&$ VARIABILITY}

Aperture photometry was performed on all pre-selected sources in each epoch of the version 1.0 GOODS-South V-band images using an aperture radius of 2.5 pixels (0.075"). All images were visually inspected to ensure consistent centering across all five epochs. The photometry yielded a light curve for each source. In order to quantify variation and pick out galaxies varying significantly above the photometric noise, we calculated the average magnitude, the standard deviation around the average (${\sigma}$), and the typical photometric error which was calculated using the following formula: 
\begin{equation}
error_{\sigma} = \sqrt{\frac{\Sigma(error_{mag})^{2}}{N}}
\end{equation}
Here, error$_{mag}$ is the formal magnitude error for a galaxy nucleus in each epoch and N is the total number of measurements (in this case N=5, for 5 epochs). Error$_{\sigma}$ is then the RMS of the magnitude errors (error$_{mag}$), essentially giving an error bar on the standard deviation (${\sigma}$). We then defined the ``significance parameter'' of each object's variability in the following way: 
\begin{equation}
Significance = \frac{\sigma}{error_{\sigma}}
\end{equation}
By dividing the standard deviation by the RMS of its errors, we compare the amount of an object's variation around its average magnitude by its typical photometric error. Thus, the significance parameter of each object's variability is simply a measure of its standard deviation normalized by the RMS of the photometric error. This significance parameter should not be interpreted in terms of statistical probabilities, since the number of points (5 epochs of data) is not enough to determine a Gaussian distribution. Nonetheless, this quantity does provide a valid measurement of the level of variability found for each galaxy in our sample. We compare the significance values against optical, IR, and X-ray properties to search for correlations between optical variability and properties at other wavelengths. Table 1 lists all of our objects and their optical, IR, and X-ray properties.

\section{DISCUSSION}

Figure 1 shows the standard deviation for each galaxy nucleus (${\sigma}$) vs. average nuclear V magnitude. The error bars are error$_{\sigma}$ for each source, which is the RMS of the photometric errors in each epoch, and the solid line is 3 $\times$ error$_{\sigma}$. Objects which have large standard deviations and relatively small values of error$_{\sigma}$ clearly stand out above the solid line. We have chosen a significance value of 3 as our variability threshold, although throughout the paper and figures we carry along the significance value for each source. A significance of 3 or greater means that the standard deviation is at least three times the amount of change in the photometric errors over the five epochs. Figure 2 shows variability significance vs. average nuclear V magnitude for all objects. The solid line at a significance value of 3 is the same as the 3 $\times$ error$_{\sigma}$ line in Figure 1. X-ray sources are shown as black triangles, IR sources are shown as gray squares, and objects that appear in both catalogs are shown as asterisks. 

\begin{figure}[h]
\plotone{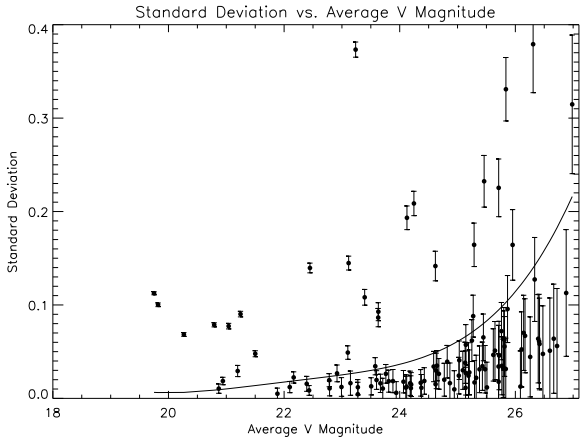}
\caption{Standard deviation vs. average V magnitude. The error bars are values of error$_{\sigma}$, the RMS of the photometric errors over five epochs. The line traces out values of three times the RMS of the photometric errors at each magnitude. Objects with standard deviations greater than this value are considered to be varying in the optical. \label{fig1}}
\end{figure}

We find a total of 29 variables out of the 112 galaxies selected via X-ray and mid-IR emission, resulting in a 26$\%$ variability rate for our candidates. Eight of the 14 objects (57$\%$) which are both mid-IR power law-selected AGN candidates and X-ray sources show significant optical variability. Many of these are also among the brightest optical sources. Detection in both X-rays and the MIR indicates the presence of an AGN with some dust present to reprocess a portion of the light and re-emit it in the infrared. Among the other variable objects, those detected only in X-rays are likely AGN enshrouded by little dust, resulting in no significant reprocessed light and were thus not identified in the mid-IR-selected sample. Two of the objects detected as optical variables were identified via mid-IR power-law SEDs only and do not show X-ray emission (squares). These sources are likely obscured AGNs that do not show up in X-rays because much of the X-ray, UV, and optical light is reprocessed into the mid-IR. Thus, while these objects would not be selected as AGN via X-ray emission, they do show optical and IR evidence of their AGN nature. In total, we find that 10 out of 22 mid-IR-selected AGN candidates (45$\%$) and 27 out of 104 X-ray-selected AGN candidates (26$\%$) show optical variability over the course of 6 months. 

\begin{figure}[h]
\plotone{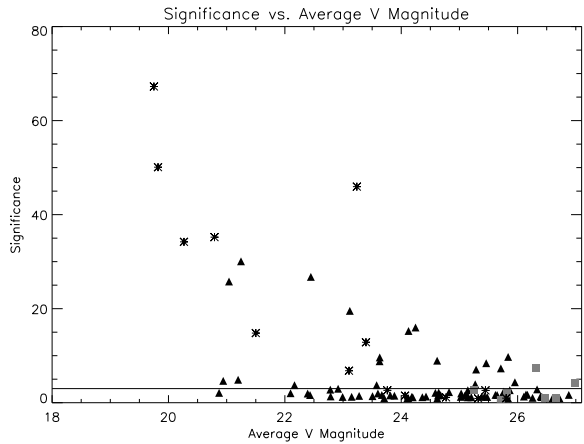}
\caption{Significance vs. average V magnitude (r=2.5 pixels) for all objects in this study. Black triangles represent objects detected only in X-rays, gray squares represent objects selected via IR power law behavior, and asterisks represent objects that appear in both catalogs. The solid line at significance 3 separates the variable objects from the nonvariable objects as discussed in Section 4. 
\label{fig2}}
\end{figure}

We further explore the properties of the mid-IR-selected AGN in Figure 3. Alonso-Herrero et al. (2006) separate BLAGN-like SEDs from NLAGN-like SEDs at $\alpha$ $\backsimeq$ -0.9, where steeper (i.e., more negative) values represent NLAGNs and shallower power law SEDs are classified as BLAGNs. This figure reveals that 7 out of 11 galaxies classified as BLAGN are optical variables (64$\%$), 2 out of 4 borderline sources are variable (50$\%$), and only one of 7 NLAGNs shows optical variability (14$\%$). Traditionally we expect that optical variability should be most observable among Type 1 AGN, since the leading theory for the cause of variability is based on instabilities within the accretion disk \citep{per06}. These instabilities would be most apparent in Type 1 AGN, which give us a clear view of this part of the system in the unified model. Therefore, our findings are consistent with this expectation, but also show that many borderline and a small fraction of NLAGNs can also be identified as optical variables.

\begin{figure}[h]
\plotone{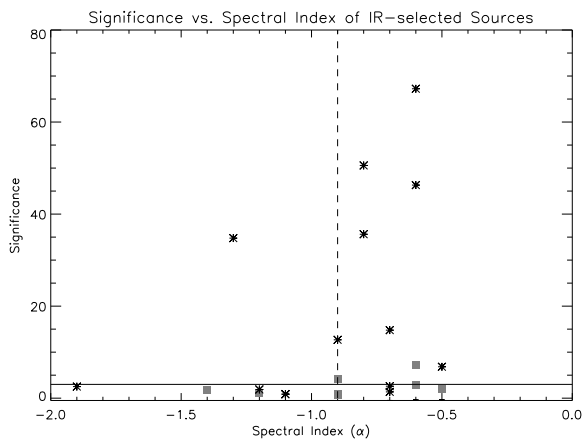}
\caption{Optical variability significance vs. spectral index for mid-IR-selected galaxies. Gray squares represent those sources found only via mid-IR selection and asterisks represent objects identified as both IR and X-ray sources. The vertical line at $\alpha$ = -0.9 shows the approximate dividing line between SEDs classified as NLAGNs (steeper SEDs having more negative spectral indices) and BLAGNs, and the solid horizontal line at significance 3 separates variables from non-variables.
\label{fig3}}
\end{figure}

Of the X-ray sources in our study, only 26$\%$ are found to be optical variables. To interpret these statistics, we first calculated the X-ray-to-optical flux ratios for the X-ray sources to determine that they are indeed AGN. We use the R-band magnitudes for the X-ray sources from Giacconi et al. (2002), taken with the FORS1 camera at the VLT and the Wide Field Imager (WFI) on the ESO-MPG 2.2 meter telescope at La Silla. There are 93 objects with published R magnitudes in our sample of 104 X-ray objects (89$\%$). We find that the X-ray sources in our survey cluster around log(F$_{x}$/F$_{opt}$)=0 and cover the range between log(F$_{x}$/F$_{opt}$)=1 to -2, which is the general range of values for AGN \citep{com02,szo04}. We also find four sources with log(F$_{x}$/F$_{opt}$)$<$-2 (AID 182, 189, 192, and 207). These sources have optical variability signficance values of 0.83, 4.88, 2.09, and 4.64, respectively. The two with high significance values ($>$3) may be AGN which simply have low F$_{x}$/F$_{opt}$ values, consistent with some optical variables identified in the HDF-N \citep{sgk03}.  The other two, having both low variability significance and low F$_{x}$/F$_{opt}$, may in fact not be AGN. Thus, 98$\%$ of the X-ray sources have F$_{x}$/F$_{opt}$ values consistent with AGN. 

To investigate trends among variability and X-ray band ratio, we used the published values of Alexander et al. (2003), where the band ratio (BR) is the ratio of the hard X-ray band (2-8 keV; HB) source counts to the soft X-ray band (0.5-2 keV; SB) counts. We arbitrarily separate ``hard" sources from ``soft" sources at a band ratio of BR=0.5, with soft sources having BR$<$0.5. Of the 100 X-ray-selected AGN candidates for which a band ratio is measured, 37$\%$ are soft sources and 63$\%$ are hard sources. Figure 4 shows variability significance vs. X-ray band ratio. We find that 19 out of the 37 X-ray sources with soft band ratios are variable (51$\%$). There are very few variable objects with band ratios greater than 0.5 and there are no variables among those sources with band ratios harder than 2. We find only 8 variables out of 62 sources with a band ratio greater than 0.5. These findings are consistent with the expectation that harder band ratios indicate a more obscured source, which will likely show less optical variability as dust may be obscuring and reprocessing the optical AGN light. Objects which exhibit soft band ratios are less obscured and therefore much easier to detect as variables. In summary, 51$\%$ of soft X-ray sources (BR$<$0.5) are optical variables, 16$\%$ of sources having band ratios in the range 0.5$<$BR$<$2 are variables (8 of 48), and no objects with BR$>$2 are detected as optical variables. We clearly observe a decreasing level of optical variability with increasing X-ray hardness. 

\begin{figure}[h]
\plotone{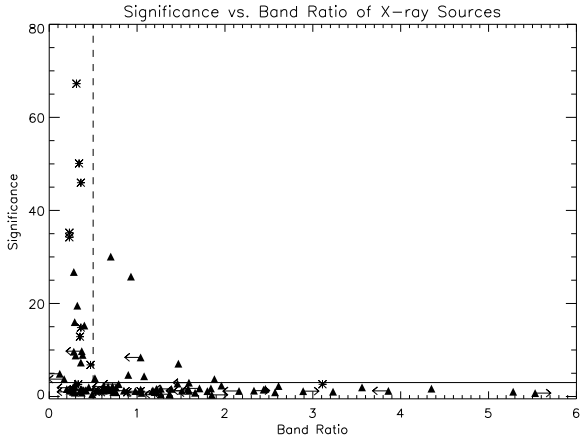}
\caption{Significance vs. X-ray band ratio, defined as BR=HB/SB. Black triangles represent sources detected only in X-rays, while asterisks are sources detected in both X-rays and selected in the mid-IR. Leftward-pointing arrows are objects for which the band ratio is only an upper limit; rightward-pointing arrows are objects for which it is a lower limit. The vertical line at 0.5 separates hard (BR$>$0.5) and soft (BR$<$0.5) band ratios, and the solid horizontal line at significance 3 separates variables from non-variables. 
\label{fig4}}
\end{figure}

Webb $\&$ Malkan (2000) find that the amplitude of optical variability in AGN increases on timescales of 1-100 days, with 60$\%$ of AGNs varying on month-to-month timescales. The objects in their study are estimated to have magnitudes similar to the objects in our sample. The results from Figure 4 are consistent with this expectation, as we see 51$\%$ of objects with soft band ratios showing variability. The 49$\%$ of soft sources that do not show variability are thus likely to be AGN that do not have variability timescales to which our survey is sensitive - our baseline of months is not sufficient to pick up variability in these sources. Webb $\&$ Malkan also found that the observed variability does not appear to depend on other AGN properties; a similar study over the same timescale of the same galaxies would reveal the same percentage of variability, though the specific galaxies found to show the greatest variability would change. Thus, among the soft X-ray sources, it is plausible that all would show optical variability if observed over a longer time baseline (i.e., several years). This interpretation is also valid for the mid-IR-selected sources. We found that 50 to 63$\%$ of the BLAGN/borderline SEDs are optical variables. Again, all may be variable when observed over longer time baselines. The drop in the fraction of optical variables among harder X-ray sources and the NLAGN SEDs among IR-selected sources, however, is likely due to increased levels of obscuring material in these types of AGN.

For the many non-variable X-ray sources, a coexisting possibility is that some fall into the class of X-ray Bright Optically Normal Galaxies (XBONGs), objects which appear as AGNs via X-ray flux but lack optical evidence for accretion (Comastri 2002). Such objects tend to have hard X-ray flux ratios indicative of obscured accretion. Some of these objects can be explained as AGN which are dominated by their host galaxy light, which would dilute the AGN light and mask optical evidence for accretion. Additionally, Rigby et al. (2006) studied hard X-ray-selected AGNs in host galaxies having a wide range of inclination angles. They concluded that the optical dullness in some AGN within host galaxies that are not face-on or spheroidal may be a result of obscuring material aligned with the host galaxy and far from the ionizing nuclear activity. This is consistent with our finding of fewer optical variables among harder X-ray sources.

Finally, we have investigated whether there is any relationship among the published X-ray band ratios and the multiwavelength SED classifications of Alonso-Herrero et al. (2006), and how that might relate to the detection of optical variability. Interestingly, of those sources selected via both X-ray and mid-IR criteria, most (10 out of 14) have soft band ratios. Somewhat surprisingly, the hardest source from this subset of galaxies (AID 179) is classified with a BLAGN SED. All of the optical variables detected in both X-rays and via mid-IR selection are soft X-ray sources with BR$<$0.5. Thus, we do not observe a clear correlation between the X-ray and multiwavelength SED AGN classifications. Similarly, Barmby et al. (2006) finds only marginal agreement between AGN classification determined separately based on mid-IR SEDs and X-ray band ratios, although the amount of obscuration present should in principle affect both wavelength regimes. They list variations in the gas-to-dust ratio or a range of intrinsic AGN properties as possible reasons that the IR and X-ray classifications do not tend to agree. Such reasons may also explain the many non-X-ray-detected galaxies with IR properties similar to those that are detected in X-ray surveys.

\section{CONCLUSIONS}
 
We have investigated the presence of optical variability in AGN candidates selected via X-ray emission and mid-IR power law SEDs. We find a range of variability amplitudes among these candidates, with 26$\%$ having optical flux variations around the mean that are at least 3 times the typical photometric error. The most significant variables are both mid-IR- and X-ray-selected AGN and are bright optical sources. Most of the variables, however, are X-ray-only selected sources with soft band ratios, indicating a relatively unobscured AGN with negligible amounts of dust near the ionizing source. These findings are consistent with the expectation that optical variability selects primarily Type 1, unobscured AGN.  

We find several indications, however, that optical variability is also observable among more obscured AGN, albeit with a much lower detection rate. Eight of the optically variable X-ray sources in our survey have band ratios greater than 0.5. A higher band ratio indicates higher column densities of obscuring material along the line of sight. We find that the optical variability significance decreases with increasing amounts of obscuration. Using the multiwavelength
SED classifications of Alonso-Herrero et al. (2006), we find that while most (70$\%$) of the optical variables are classified as BLAGNs, 20$\%$ are borderline NLAGN/BLAGN SEDs and 10$\%$ (1 of 10) have a NLAGN SED. Finally, we detected optical variability for two mid-IR power law-selected AGN that are not detected in X-rays. Such sources may be heavily obscured AGN, where X-rays and much of the optical/UV light is blocked by dust which re-emits the light in the mid-IR. These sources are among the faintest optical sources in our survey and also lie close to the variability threshold.

A large fraction of the AGN candidates (74$\%$), however, do not show optical variability significant enough to produce flux changes greater than 3 times the typical photometric error. Among the relatively unobscured sources (i.e., those detected in the X-ray with band ratios less than 0.5), about 50$\%$ are not variable. This is consistent with the expectation that only $\sim$60$\%$ of AGN show optical variability on month-to-month timescales \citep{web00}. Thus, the majority of unobscured AGN that do not show optical variability may be optical variables when observed on longer time intervals of years. The inclusion of additional epochs of imaging data for GOODS-S will help to answer this question. Another possibility is that some of the sources are in the class of ``optically dull" galaxies which exhibit AGN X-ray luminosities but show no optical evidence for accretion. Some may be dominated by light from the non-varying host galaxy, while others may reside in galaxies with large amounts of obsuring material in the host.  

\acknowledgments

This paper is based on observations with the NASA/ESA Hubble Space Telescope, obtained at the Space Telescope Science Institute, which is operated by the Association of Universities for Research in Astronomy, Inc., under NASA contract NAS5-26555 and observations with the Keck  telescope, made possible by the W. M. Keck Foundation and NASA. Funding was provided by STScI grant AR-09948.01-A and NSF CAREER grant 0346691.

{\it Facilities:} \facility{HST (ACS)}, \facility{Spitzer (MIPS)}, \facility{CXO (ACIS-I)}.

\clearpage

\begin{deluxetable}{cccllccccccr}
\tabletypesize{\scriptsize}
\rotate
\tablecaption{Optical, IR, $\&$ X-ray properties of all sources}\label{tbl-1}
\tablewidth{0pt}
\setlength{\tabcolsep}{0.02in}
\tablehead{
\colhead{AID\tablenotemark{a}} & \colhead{MIPSID\tablenotemark{b}} & \colhead{GOODS ID} & \colhead{RA} & \colhead{DEC} & \colhead{$<$V$_{nuc}$$>$} & \colhead{$\sigma$} & \colhead{Error$_{\sigma}$} &
\colhead{Significance} & \colhead{$\alpha$\tablenotemark{c}} & \colhead{$\Delta$$\alpha$} & \colhead{Band Ratio\tablenotemark{d}}}

\startdata

113\tablenotemark{1} &  -	   & J033217.14-275402.5  & 53.0714194 & -27.9007025  & 25.83  & 0.03 & 0.03 &  0.95  &     -   &  -   &   1.22    \\
-   &  mips003133  & J033246.84-275121.2  & 53.1951520 & -27.8558940  & 25.25  & 0.06 & 0.02 &  2.76  &  -0.6   &  0.2 &   -    \\
-   &  mips003149  & J033214.55-275256.6  & 53.0606141 & -27.8823767  & 25.71  & 0.02 & 0.03 &  0.61  &  -0.9   &  0.2 &   -    \\
6a  &  -	   & J033216.16-274941.7  & 53.0673266 & -27.8282610  & 24.63  & 0.03 & 0.02 &  1.55  &     -   &  -   &  $<$ 0.37   \\
-   &  mips003485  & J033225.24-275226.6  & 53.1051776 & -27.8740506  & 25.81  & 0.06 & 0.03 &  1.73  &  -1.4   &  0.2 &   -    \\
-   &  mips003528  & J033234.46-275005.0  & 53.1436200 & -27.8347720  & 26.48  & 0.05 & 0.05 &  0.93  &  -0.9   &  0.2 &   -    \\
-   &  mips003618  & J033235.71-274916.0  & 53.1487963 & -27.8211171  & 26.66  & 0.06 & 0.06 &  1.09  &  -1.2   &  0.2 &   -    \\
9a  &  -	   & J033218.45-274555.9  & 53.0768751 & -27.7655258  & 24.94  & 0.01 & 0.02 &  0.48  &     -   &  -   &  $<$ 1.27   \\
-   &  mips003537  & J033240.75-274926.5  & 53.1699410 & -27.8240740  & 26.31  & 0.38 & 0.05 &  7.31  &  -0.6   &  0.2 &   -    \\
-   &  mips003871  & J033233.02-274200.4  & 53.1375810 & -27.7001130  & 25.81  & 0.06 & 0.03 &  2.02  &  -0.5   &  0.2 &   -    \\
-   &  mips010818  & J033227.19-274051.4  & 53.1133032 & -27.6809373  & 26.98  & 0.31 & 0.07 &  4.24  &  -0.9   &  0.2 &   -    \\
39  &  -	   & J033201.58-274327.0  & 53.0065900 & -27.7241626  & 24.43  & 0.02 & 0.01 &  1.31  &     -   &  -   &  0.42   \\
43  &  -	   & J033203.04-274450.1  & 53.0126503 & -27.7472400  & 26.33  & 0.13 & 0.04 &  2.83  &     -   &  -   &  0.30   \\
44  &  -	   & J033203.65-274603.7  & 53.0152257 & -27.7676828  & 23.60  & 0.02 & 0.01 &  1.96  &     -   &  -   &  3.56   \\
53  &  -	   & J033206.27-274536.7  & 53.0261253 & -27.7601830  & 24.59  & 0.03 & 0.02 &  2.10  &     -   &  -   &  $<$ 0.67   \\
60  &  mips004281  & J033207.98-274239.5  & 53.0332562 & -27.7109593  & 25.32  & 0.02 & 0.02 &  0.91  &  -1.1   &  0.2 &  0.89   \\
62  &  -	   & J033208.00-274657.3  & 53.0333350 & -27.7825706  & 26.09  & 0.01 & 0.04 &  0.33  &     -   &  -   &  0.51   \\
63  &  -	   & J033208.27-274153.5  & 53.0344441 & -27.6982065  & 25.84  & 0.33 & 0.03 &  9.74  &     -   &  -   &  $<$ 0.37   \\
65  &  -	   & J033208.53-274648.3  & 53.0355495 & -27.7800797  & 23.28  & 0.00 & 0.01 &  0.35  &     -   &  -   &  $>$ 1.85   \\
66  &  mips013611  & J033208.66-274734.4  & 53.0360946 & -27.7928810  & 19.82  & 0.10 & 0.00 &  50.11 &  -0.8   &  0.2 &  0.34   \\
68  &  -	   & J033209.45-274806.8  & 53.0393617 & -27.8018870  & 21.04  & 0.08 & 0.00 &  25.78 &     -   &  -   &  0.93   \\
73  &  -	   & J033210.76-274234.6  & 53.0448370 & -27.7096055  & 22.99  & 0.01 & 0.01 &  1.22  &     -   &  -   &  $<$ 1.59   \\
73  &  -	   & J033210.76-274234.6  & 53.0448370 & -27.7096055  & 22.92  & 0.03 & 0.01 &  2.97  &     -   &  -   &  $<$ 1.59   \\
75  &  -	   & J033210.91-274343.1  & 53.0454675 & -27.7286273  & 25.65  & 0.05 & 0.03 &  1.64  &     -   &  -   &  $<$ 1.39   \\
76  &  mips004258  & J033210.91-274414.9  & 53.0454704 & -27.7374846  & 23.39  & 0.11 & 0.01 &  12.86 &  -0.9   &  0.2 &  0.35   \\
80  &  -	   & J033211.41-274650.0  & 53.0475311 & -27.7805481  & 24.36  & 0.02 & 0.01 &  1.17  &     -   &  -   &  0.98   \\
83  &  -	   & J033212.20-274530.1  & 53.0508213 & -27.7583569  & 24.18  & 0.02 & 0.01 &  1.24  &     -   &  -   &  $<$ 1.51   \\
84  &  -	   & J033212.22-274620.6  & 53.0509337 & -27.7724017  & 25.51  & 0.01 & 0.03 &  0.43  &     -   &  -   &  1.37   \\
88  &  -	   & J033213.25-274240.9  & 53.0551915 & -27.7113507  & 22.43  & 0.01 & 0.01 &  1.63  &     -   &  -   &  1.22   \\
91  &  -	   & J033213.85-274248.9  & 53.0577260 & -27.7135792  & 25.72  & 0.05 & 0.04 &  1.25  &     -   &  -   &  $>$ 2.33   \\
94  &  -	   & J033214.00-275100.7  & 53.0583525 & -27.8502044  & 25.13  & 0.04 & 0.02 &  1.73  &     -   &  -   &  1.84   \\
95  &  -	   & J033214.08-274230.4  & 53.0586652 & -27.7084382  & 25.46  & 0.23 & 0.03 &  8.40  &     -   &  -   &  $<$ 1.04   \\
96  &  -	   & J033214.43-275110.8  & 53.0601151 & -27.8530055  & 24.82  & 0.04 & 0.02 &  2.25  &     -   &  -   &  2.61   \\
97  &  -	   & J033214.45-274456.6  & 53.0601915 & -27.7490538  & 26.42  & 0.06 & 0.05 &  1.19  &     -   &  -   &  $<$ 2.16   \\
102 &  -	   & J033214.98-274224.9  & 53.0624129 & -27.7069101  & 25.02  & 0.02 & 0.02 &  1.24  &     -   &  -   &  0.62   \\
106 &  -	   & J033215.80-275324.7  & 53.0658257 & -27.8902004  & 25.80  & 0.06 & 0.03 &  1.89  &     -   &  -   &  $<$ 0.27   \\
107 &  -	   & J033215.79-274629.7  & 53.0657934 & -27.7749284  & 24.62  & 0.02 & 0.02 &  0.94  &     -   &  -   &  -     \\
113 &  -	   & J033216.74-274327.5  & 53.0697345 & -27.7243116  & 24.11  & 0.01 & 0.01 &  0.94  &     -   &  -   &  $<$ 1.20   \\
114 &  -	   & J033216.85-275007.5  & 53.0702116 & -27.8354160  & 26.14  & 0.07 & 0.04 &  1.74  &     -   &  -   &  $<$ 1.71   \\
115 &  -	   & J033217.06-274921.9  & 53.0710682 & -27.8227401  & 22.78  & 0.02 & 0.01 &  2.78  &     -   &  -   &  1.46   \\
117 &  -	   & J033217.14-274303.3  & 53.0714326 & -27.7175864  & 23.12  & 0.14 & 0.01 &  19.5  &     -   &  -   &  0.32   \\
121 &  -	   & J033218.01-274718.5  & 53.0750617 & -27.7884859  & 24.12  & 0.01 & 0.01 &  1.00  &     -   &  -   &  $<$ 0.27   \\
122 &  -	   & J033218.24-275241.4  & 53.0760024 & -27.8781606  & 24.62  & 0.14 & 0.02 &  8.94  &     -   &  -   &  0.38   \\
123 &  -	   & J033218.34-275055.2  & 53.0764119 & -27.8486599  & 25.79  & 0.03 & 0.03 &  1.03  &     -   &  -   &  5.28   \\
125 &  -	   & J033218.83-275135.5  & 53.0784670 & -27.8598545  & 26.60  & 0.05 & 0.06 &  0.90  &     -   &  -   &  2.57   \\
126 &  -	   & J033219.00-274755.5  & 53.0791461 & -27.7987408  & 25.77  & 0.03 & 0.03 &  1.12  &     -   &  -   &  1.17   \\
129 &  -	   & J033219.81-274122.7  & 53.0825338 & -27.6896504  & 22.39  & 0.02 & 0.01 &  1.95  &     -   &  -   &  $<$ 0.76   \\
131 &  -	   & J033220.05-274447.2  & 53.0835448 & -27.7464484  & 26.17  & 0.07 & 0.04 &  1.69  &     -   &  -   &  4.35   \\
132 &  -	   & J033220.31-274554.7  & 53.0846122 & -27.7651813  & 25.19  & 0.02 & 0.04 &  0.69  &     -   &  -   &  -     \\
134 &  -	   & J033220.48-274732.3  & 53.0853209 & -27.7923093  & 25.20  & 0.03 & 0.02 &  1.16  &     -   &  -   &  1.80   \\
137 &  -	   & J033221.42-274231.2  & 53.0892636 & -27.7086599  & 25.14  & 0.03 & 0.02 &  1.15  &     -   &  -   &  $>$ 2.89   \\
144 &  -	   & J033222.54-274804.3  & 53.0939086 & -27.8011896  & 26.40  & 0.06 & 0.05 &  1.35  &     -   &  -   &  $<$ 0.39   \\
145 &  -	   & J033222.54-274603.8  & 53.0939227 & -27.7677328  & 25.76  & 0.07 & 0.03 &  2.35  &     -   &  -   &  1.96   \\
146 &  -	   & J033222.55-274949.8  & 53.0939514 & -27.8305093  & 26.11  & 0.05 & 0.04 &  1.30  &     -   &  -   &  1.18   \\
148 &  -	   & J033222.58-274425.8  & 53.0941038 & -27.7405110  & 24.87  & 0.02 & 0.02 &  0.75  &     -   &  -   &  -     \\
149 &  -	   & J033222.76-275224.0  & 53.0948392 & -27.8733220  & 22.10  & 0.01 & 0.01 &  2.02  &     -   &  -   &  0.62   \\
155 &  -	   & J033224.26-274126.4  & 53.1010655 & -27.6906716  & 24.19  & 0.01 & 0.01 &  0.83  &     -   &  -   &  0.37   \\
158 &  -	   & J033224.86-274706.4  & 53.1035834 & -27.7851064  & 23.71  & 0.01 & 0.01 &  0.96  &     -   &  -   &  $<$ 0.73   \\
160 &  -	   & J033224.96-275008.0  & 53.1039901 & -27.8355688  & 26.26  & 0.04 & 0.04 &  1.04  &     -   &  -   &  3.23   \\
161 &  -	   & J033224.98-274101.5  & 53.1040892 & -27.6837528  & 23.67  & 0.02 & 0.01 &  1.64  &     -   &  -   &  2.46   \\
162 &  -	   & J033225.11-275043.3  & 53.1046078 & -27.8453489  & 25.09  & 0.03 & 0.02 &  1.47  &     -   &  -   &  2.44   \\
163 &  mips010787  & J033225.17-274218.8  & 53.1048578 & -27.7052197  & 24.07  & 0.02 & 0.01 &  1.49  &  -0.7   &  0.2 &  0.28   \\
164 &  -	   & J033225.16-275450.1  & 53.1048319 & -27.9139259  & 25.29  & 0.02 & 0.02 &  0.73  &     -   &  -   &  $>$ 5.53   \\
167 &  -	   & J033225.74-274936.4  & 53.1072675 & -27.8267670  & 23.94  & 0.01 & 0.01 &  0.50  &     -   &  -   &  $<$ 1.37   \\
173 &  mips013087  & J033226.49-274035.5  & 53.1103938 & -27.6765399  & 20.27  & 0.07 & 0.00 &  34.21 &  -1.3   &  0.2 &  0.23   \\
174 &  mips010823  & J033226.67-274013.4  & 53.1111149 & -27.6703838  & 25.80  & 0.03 & 0.03 &  0.97  &  -1.1   &  0.2 &  0.53   \\
177 &  mips000309  & J033227.01-274105.0  & 53.1125287 & -27.6847238  & 19.75  & 0.11 & 0.00 &  67.25 &  -0.6   &  0.2 &  0.31   \\
179 &  mips003888  & J033227.62-274144.9  & 53.1150981 & -27.6958053  & 23.76  & 0.03 & 0.01 &  2.63  &  -0.7   &  0.2 &  3.11   \\
181 &  -	   & J033228.74-274620.4  & 53.1197554 & -27.7723316  & 24.68  & 0.03 & 0.02 &  1.60  &     -   &  -   &  0.20   \\
182 &  -	   & J033228.81-274355.6  & 53.1200615 & -27.7321250  & 21.88  & 0.00 & 0.01 &  0.83  &     -   &  -   &  0.29   \\
188 &  -	   & J033229.85-275105.9  & 53.1243696 & -27.8516331  & 25.62  & 0.05 & 0.03 &  1.66  &     -   &  -   &  1.58   \\
189 &  -	   & J033229.88-274424.4  & 53.1244949 & -27.7401248  & 21.20  & 0.03 & 0.01 &  4.88  &     -   &  -   &  $<$ 0.12   \\
191 &  mips014635  & J033229.98-274529.9  & 53.1249148 & -27.7583013  & 21.50  & 0.05 & 0.00 &  14.81 &  -0.7   &  0.2 &  0.36   \\
192 &  -	   & J033229.99-274404.8  & 53.1249588 & -27.7346753  & 20.87  & 0.01 & 0.01 &  2.09  &     -   &  -   &  0.29   \\
193 &  -	   & J033230.06-274523.5  & 53.1252547 & -27.7565350  & 23.63  & 0.09 & 0.01 &  9.63  &     -   &  -   &  0.28   \\
195 &  -	   & J033230.22-274504.6  & 53.1258995 & -27.7512749  & 22.45  & 0.14 & 0.01 &  26.77 &     -   &  -   &  0.28   \\
196 &  -	   & J033231.36-274725.0  & 53.1306522 & -27.7902710  & 25.71  & 0.03 & 0.04 &  0.77  &     -   &  -   &  $<$ 1.05   \\
197 &  mips003938  & J033231.46-274623.2  & 53.1310696 & -27.7731060  & 24.77  & 0.02 & 0.02 &  1.13  &  -1.2   &  0.2 &  $<$ 1.04   \\
199 &  -	   & J033232.04-274451.7  & 53.1335033 & -27.7477042  & 25.14  & 0.06 & 0.02 &  2.65  &     -   &  -   &  -     \\
203 &  -	   & J033232.99-274117.0  & 53.1374396 & -27.6880584  & 26.41  & 0.06 & 0.05 &  1.21  &     -   &  -   &  $<$ 3.86   \\
207 &  -	   & J033233.46-274312.8  & 53.1394133 & -27.7202143  & 20.94  & 0.02 & 0.00 &  4.64  &     -   &  -   &  0.90   \\
212 &  -	   & J033234.34-274350.1  & 53.1430925 & -27.7305813  & 24.64  & 0.03 & 0.02 &  2.16  &     -   &  -   &  0.72   \\
214 &  -	   & J033234.95-275511.2  & 53.1456346 & -27.9197732  & 25.14  & 0.01 & 0.02 &  0.51  &     -   &  -   &  0.49   \\
215 &  -	   & J033235.04-274932.6  & 53.1459896 & -27.8257333  & 26.72  & 0.06 & 0.06 &  0.91  &     -   &  -   &  $<$ 0.76   \\
216 &  -	   & J033235.10-274410.6  & 53.1462691 & -27.7362726  & 25.43  & 0.03 & 0.03 &  1.33  &     -   &  -   &  $<$ 0.70   \\
227 &  -	   & J033236.72-274406.4  & 53.1529803 & -27.7351237  & 25.03  & 0.04 & 0.02 &  1.98  &     -   &  -   &  0.45   \\
229 &  -	   & J033237.46-274000.1  & 53.1560803 & -27.6666919  & 23.63  & 0.09 & 0.01 &  8.82  &     -   &  -   &  0.30   \\
230 &  mips010697  & J033237.76-275212.3  & 53.1573435 & -27.8700853  & 25.45  & 0.07 & 0.03 &  2.61  &  -1.9   &  0.2 &  0.33   \\
234 &  mips003915  & J033238.12-273944.8  & 53.1588307 & -27.6624444  & 20.79  & 0.08 & 0.00 &  35.21 &  -0.8   &  0.2 &  0.23   \\
236 &  -	   & J033238.87-274733.2  & 53.1619461 & -27.7925487  & 25.87  & 0.10 & 0.04 &  2.67  &     -   &  -   &  $<$ 0.79   \\
237 &  -	   & J033238.76-275121.6  & 53.1615043 & -27.8560057  & 26.88  & 0.11 & 0.07 &  1.66  &     -   &  -   &  1.27   \\
242 &  -	   & J033239.09-274601.8  & 53.1628593 & -27.7671602  & 21.25  & 0.09 & 0.00 &  30.07 &     -   &  -   &  0.70   \\
244 &  -	   & J033239.46-275031.8  & 53.1644211 & -27.8421693  & 24.20  & 0.01 & 0.01 &  1.15  &     -   &  -   &  0.35   \\
248 &  -	   & J033241.40-274717.1  & 53.1725205 & -27.7880947  & 24.37  & 0.01 & 0.01 &  0.80  &     -   &  -   &  $<$ 1.66   \\
250 &  -	   & J033241.87-274359.9  & 53.1744526 & -27.7332996  & 25.71  & 0.23 & 0.03 &  7.30  &     -   &  -   &  0.36   \\
251 &  mips003108  & J033241.85-275202.5  & 53.1743886 & -27.8673533  & 23.10  & 0.05 & 0.01 &  6.80  &  -0.5   &  0.2 &  0.47   \\
254 &  -	   & J033242.86-274702.7  & 53.1785932 & -27.7840820  & 25.27  & 0.09 & 0.02 &  3.93  &     -   &  -   &  0.52   \\
256 &  mips009708  & J033243.24-274914.2  & 53.1801493 & -27.8206046  & 23.24  & 0.37 & 0.01 &  45.95 &  -0.6   &  0.2 &  0.36   \\
258 &  -	   & J033244.01-274635.0  & 53.1833644 & -27.7763817  & 25.38  & 0.03 & 0.03 &  1.24  &     -   &  -   &  0.23   \\
260 &  -	   & J033244.27-275141.1  & 53.1844471 & -27.8614196  & 22.17  & 0.02 & 0.01 &  3.74  &     -   &  -   &  $<$ 0.17   \\
261 &  -	   & J033244.31-275251.3  & 53.1846401 & -27.8809173  & 24.25  & 0.21 & 0.01 &  15.99 &     -   &  -   &  0.29   \\
262 &  -	   & J033244.44-274819.0  & 53.1851517 & -27.8052754  & 23.81  & 0.02 & 0.01 &  1.56  &     -   &  -   &  0.36   \\
263 &  -	   & J033244.45-274940.2  & 53.1852272 & -27.8278365  & 25.29  & 0.16 & 0.02 &  7.06  &     -   &  -   &  1.47   \\
264 &  -	   & J033244.60-274835.9  & 53.1858335 & -27.8099684  & 25.48  & 0.03 & 0.03 &  1.23  &     -   &  -   &  0.55   \\
265 &  -	   & J033245.02-275439.6  & 53.1875770 & -27.9109950  & 23.15  & 0.02 & 0.01 &  1.25  &     -   &  -   &  $<$ 0.85   \\
266 &  -	   & J033245.11-274724.0  & 53.1879518 & -27.7899986  & 23.28  & 0.01 & 0.01 &  1.45  &     -   &  -   &  $<$ 0.66   \\
267 &  -	   & J033245.68-275534.4  & 53.1903431 & -27.9262231  & 22.79  & 0.01 & 0.01 &  1.34  &     -   &  -   &  0.41   \\
269 &  -	   & J033246.41-275414.0  & 53.1933716 & -27.9038774  & 23.88  & 0.02 & 0.01 &  1.49  &     -   &  -   &  1.04   \\
273 &  -	   & J033247.18-275147.5  & 53.1965772 & -27.8632067  & 25.95  & 0.16 & 0.04 &  4.35  &     -   &  -   &  $<$ 1.08   \\
276 &  -	   & J033248.18-275256.6  & 53.2007350 & -27.8823907  & 23.58  & 0.03 & 0.01 &  3.74  &     -   &  -   &  1.88   \\
286 &  -	   & J033252.88-275119.8  & 53.2203537 & -27.8555099  & 24.13  & 0.19 & 0.01 &  15.25 &     -   &  -   &  0.40   \\
292 &  -	   & J033256.71-275319.1  & 53.2362800 & -27.8886316  & 23.51  & 0.01 & 0.01 &  1.40  &     -   &  -   &  $<$ 0.62   \\

\enddata		
\tablenotetext{1}{This is an additional source from the Giacconi et al. (2002) catalog; the ID number listed is that given by Giacconi et al., not that of Alexander et al. (2003)} 	
\tablenotetext{a}{AID stands for Alexander ID (Alexander et al. 2003)}
\tablenotetext{b}{MIPSID is the MIPS 24 $\mu$m name (Alonso-Herrero et al. 2006)}
\tablenotetext{c}{$\alpha$ is the spectral index of IR-selected galaxies (Alonso-Herrero et al. 2006)}
\tablenotetext{d}{Band Ratio is defined as Hard Band/Soft Band (Alexander et al. 2003)}

\end{deluxetable}																	    

\end{document}